\documentclass{iopart}
\usepackage{graphicx}
\usepackage{dcolumn} 
\usepackage{bm} 
\usepackage[dvips]{color}

\begin{document}

\bibliographystyle{unsrt}
\def\ARNPS{ {\it Ann.~Rev.~Nucl.~Part.~Sci.~}}

\def\PR{{\it Phys.~Rep.~}}
\def\Preprint{{\it Preprint~}}

\title{HBT: A (mostly) experimental overview}

\author{Dan Magestro}

\address{Department of Physics, Ohio State University, Columbus, OH 43212}

\begin{abstract}
I will present a review of the field of Hanbury Brown-Twiss
interferometry in relativistic heavy-ion collisions.  The ``HBT puzzle" is explored in detail, emphasizing recent theoretical attempts to understand the persisting puzzle.  I also present recent experimental results on azimuthally sensitive HBT, HBT of direct photons, and some surprises in the comparison of HBT results from p+p and Au+Au collisions at RHIC.
\end{abstract}



\section{Introduction}
Among the large set of measurements predicted \cite{lastCall} to uncover the formation 
and nature of a quark-gluon plasma (QGP) in relativistic heavy-ion collisions, and the corresponding phase transition between
this plasma and the more familiar hadronic phase, Hanbury Brown-Twiss interferometry (HBT) \cite{hbtReviews} surely is the most maligned.  This Quark Matter conference marks the three-year anniversary of the first presentation of experimental HBT results \cite{laue} from RHIC, which showed that the ``standard" HBT source radii in Au+Au collisions at $\sqrt{s}=130$ GeV (introduced below) are quantitatively similar to previous measurements at both AGS and SPS.  The CERES Collaboration illustrated this nicely in results \cite{ceresHBT} presented at the last Quark Matter.  The lack of an energy dependence, the subsequent inconsistency with otherwise-successful hydrodynamical models, and the failure of most alternative explanations, led to the moniker ``HBT puzzle" \cite{heinzPanic02,dumitru02}.  

Whether the ``HBT puzzle" continues to be cause for concern in our field remains to be seen, as it has recently taken a back seat to studies that do show strong differences at RHIC compared to lower energies (most notably higher $p_T$ effects attributed to jet quenching \cite{jetQuench}).  The puzzle has also been somewhat cast aside as a simple lack of adequate description for the freeze-out process (most notably at this conference by Miklos Gyulassy, an early advocate of HBT as phase transition signature \cite{RischkeGyulassy}).  While it is clear that understanding the evolution of the hot, dense, expanding system requires an understanding of its space-time structure accessed by HBT studies, it is not clear to me whether this understanding can be bypassed when making QGP discovery claims based on other observables.  In that sense, the HBT puzzle still warrants much theoretical and experimental effort, perhaps more than it currently receives.

Providing a ray of optimism, Reinhard Stock, in his Quark Matter introductory overview, characterized HBT as bringing us ``from enlightenment to tragedy, and back to enlightenment, on a time scale of four years."  With these words in mind, in this overview I assess the status of HBT measurements and the persisting puzzle, mostly from an experimentalist's standpoint.  I also discuss new developments in two-particle interferometry that contribute more to our understanding of heavy-ion collisions at relativistic energies. (For an experimental summary of related results presented at this conference, see H. Appelsh\"auser's contribution to these proceedings \cite{harry}.)

\section{HBT in relativistic heavy-ion collisions}

The experimental technique of using two-particle interferometry to relate the 
{\it momentum space} separation of particles to their separation in {\it
space-time} is well-established \cite{hbtReviews}.  In the case of identical bosons, {\it e.g.}~$\pi^+$ mesons, quantum interference among the particles leads to an enhancement of
pairs with small momentum difference ${\mathbf q}$ (Bose-Einstein enhancement).  To isolate the small set of correlated pairs that undergo this quantum interference from the enormous amount of uncorrelated pairs in an event, a correlation function $C({\mathbf q})$ is formed in which pairs from real events are divided by pairs from different events.  In heavy-ion collisions, $C({\mathbf q})$ is often constructed in three dimensions and fit to a three-dimensional Gaussian:
\begin{equation}\label{eq:prattBertsch}
C({\mathbf q}) = \frac{\rm{real\ pairs}}{\rm{mixed\ pairs}} = N \bigl[ 1 + \lambda e^{ - q_{\rm out}^2 R_{\rm out}^2 - q_{\rm side}^2 R_{\rm side}^2 - q_{\rm long}^2 R_{\rm long}^2} \bigr],
\end{equation}
where the subscripts indicate the long (parallel to beam), side (perpendicular to beam and 
total pair momentum ${\mathbf k}$) and out (perpendicular to $q_l$ and $q_s$) decomposition of ${\mathbf q}$.  $N$ is a normalization constant.  The $R$'s in Eq.~\ref{eq:prattBertsch}, known as the HBT radii, quantify the widths of the Gaussians and represent the apparent size of the particle source, which may depend on the transverse momentum slice under study ({\it i.e.} homogeneity regions \cite{sinyukov}).  In practice, final-state effects such as Coulomb also contribute to $C({\mathbf q})$ and need to be accounted for.  I won't discuss these here except to note that nearly all heavy-ion studies now have adopted an improved Coulomb treatment \cite{harry}.

The purpose of HBT studies in heavy-ion collisions is to explore the space-time evolution and freeze-out of the system.  This can be thought of as three-fold: the spatial distribution of the emission points, the time length of emission, and the dynamical properties of the system as it evolves.  HBT serves as a tool for disentangling these contributions, and the out-side-long decomposition of ${\mathbf q}$ is chosen for that reason.  Experimentally, HBT radii are studied as differentially as statistics and detector configurations allow; see Table 1.  
\begin{table}
\caption{\label{tb:diff} Some of the HBT differential studies underway or undertaken recently in heavy-ion collisions.}
\begin{indented}
\item[]\begin{tabular}{@{}lll}
\br
Diff.~quantity&What it investigates&Recent results\\
\mr
Beam energy & Onset effects, transition phenomena & \cite{ceresHBT,star130} \\
Tranverse momentum & Dynamics, collective expansion & \cite{phenix130,phenix200}\\
Particle type & Hydrodynamic $m_T$ scaling & \cite{na44,na49} \\
Collision system & Origin of Bose-Einstein enhancement & \cite{tom} \\
Azimuthal angle & Spatial anisotropy, system evolution & \cite{starAsHBT} \\
\br
\end{tabular}
\end{indented}
\end{table}
At RHIC, greatly increased pion production ($dN/dy \sim 300$ per flavor \cite{dndy}) due to higher collision energies makes triple-differential HBT analyses possible, {\it e.g.}~
$k_T \times {\rm centrality} \times \phi_{\rm pair}$ \cite{starAsHBT}.

\begin{figure}[bt]
\begin{center}	
{\includegraphics*[width=4.0in]{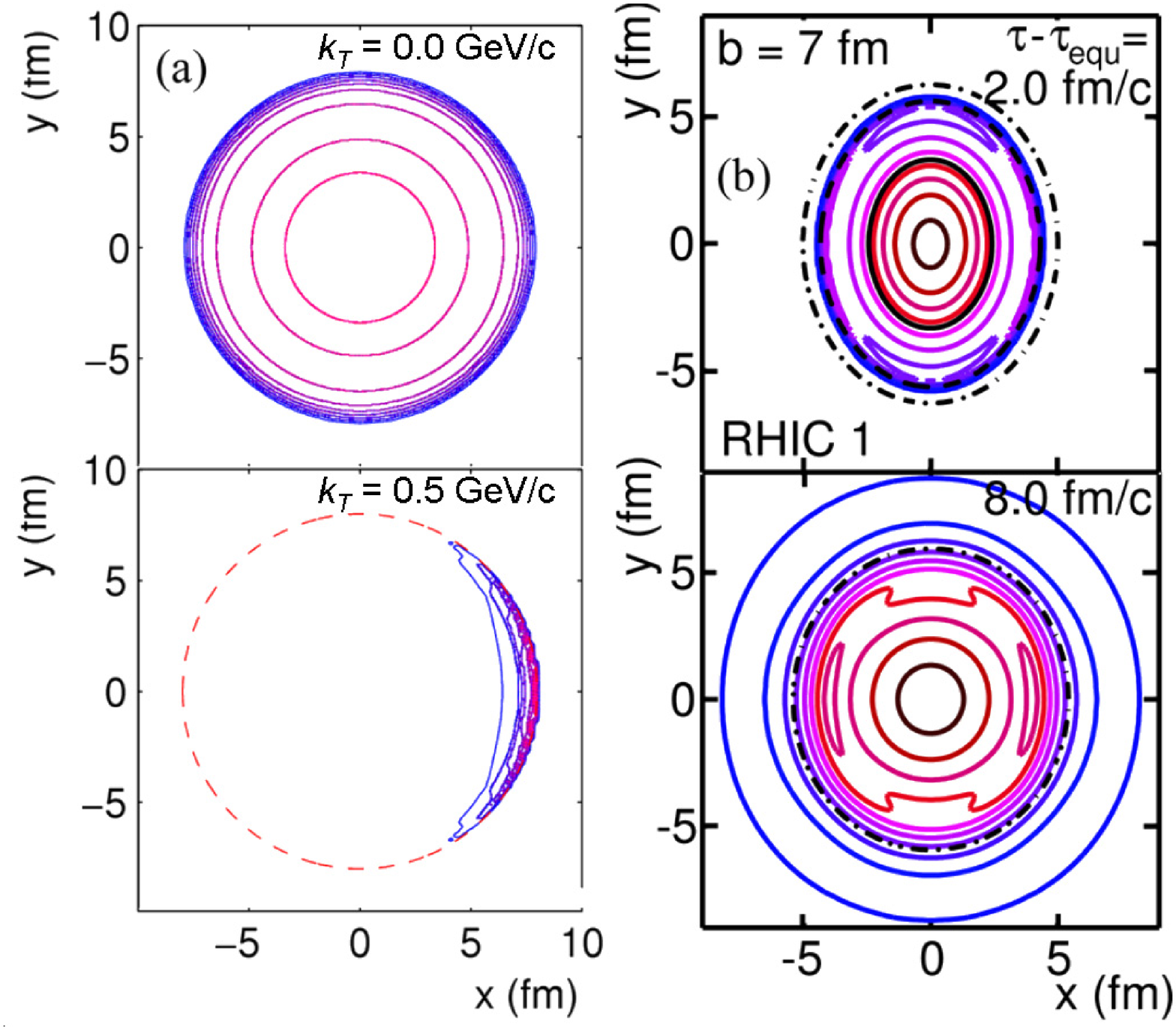}}
\end{center}
  \caption{(a) Contours representing densities of emissions points in the transverse plane for $k_T = 0.0$ GeV/c (top) and $k_T = 0.5$ GeV/c (bottom).  The emission direction is to the right, illustrating that higher $p_T$ particles emerge nearer to the surface.  (b) Contours of constant energy density at two different times in the evolution of a noncentral collision.  Both figures are hydrodynamic calculations taken from Ref.~\cite{heinzKolbQGP3}.} \label{fig:one}
\end{figure} 
Highlighting two differential studies in particular: (a) The transverse momentum ($k_T$) dependence of the HBT radii for identical pions probably is studied most often, under the model-dependent view that space-momentum correlations in the source are due mostly to collective expansion \cite{hbtReviews}.  As the source expands, radial flow pushes higher $p_T$ particles more at the surface  (Figure 1(a)).  Within this picture, analytical expressions have been derived to extract the expansion velocity and emission duration from the $m_T$ ($m_T = \sqrt{p_T^2 + m^2}$) dependence of the HBT radii. This is discussed further in section 3.  (b) HBT studies relative to the reaction plane in non-central collisions allow the possibility to compare the expanded system's transverse eccentricity at freeze-out to its initial eccentricity from a nuclear overlap model calculation (Figure 1(b)).  This is discussed further in section 4.

\section{The HBT Puzzle}

The hydrodynamical approach to understanding HBT is motivated at RHIC by the model's demonstrated ability to describe soft $p_T$ spectra and elliptic flow consistently for several particle species \cite{heinzKolbQGP3}.  ``Hydro" calculations for these observables point to fast thermalization in a partonic phase, followed by hydrodynamic expansion for $\sim 15$ fm/$c$ with an intermediate phase transition. However, these calculations yield strong disagreement with HBT radii \cite{heinzKolbWW02}: $R_{\rm out}$ and $R_{\rm long}$ are overpredicted by as much as a factor 2, and $R_{\rm side}$ is somewhat underpredicted.  In particular, the measured $k_T$ dependence of $R_{\rm side}$ is in contrast to hydro and other models that predict little (if any) $k_T$ dependence.  This disagreement, and the lack of energy dependence of the HBT radii for a fixed $k_T$ bin \cite{star130}, is known as the ``HBT Puzzle."  Here I will summarize recent theoretical approaches to resolving the puzzle, illustrating along the way why the data are indeed puzzling.

The first question is whether the hydro calculation itself can be altered to agree better with the HBT radii.  This was explored in great detail by Heinz and Kolb \cite{heinzKolbQGP3}.  A summary of their findings:
\begin{itemize}
\item {\it Default initial conditions}, such that $p_T$ spectra and elliptic
flow are well-described, yield the disagreement discussed above.
\item {\it Freeze-out directly at hadronization} brings $R_{\rm out}$ and $R_{\rm long}$ close to the data, but at the cost of large disagreement with $p_T$ spectra.  $R_{\rm side}$ doesn't move, still lacking $k_T$ dependence.
\item {\it Faster thermalization or non-zero initial flow} also reduces $R_{\rm out}$ and $R_{\rm long}$ compared to default conditions, but not enough to agree with measurement.
\end{itemize}
Hirano and Tsuda \cite{hiranoTsuda} also checked the effect of maintaining proper particle abundances with chemical potentials in the hadronic phase of hydrodynamic evolution.  They found better agreement for $R_{\rm long}$ and $R_{\rm out}$, but $R_{\rm side}$ disagreed more. 

Whether or not hydro can be made to reproduce HBT radii while maintaining the strong agreement with momentum-space quantities might come down to the assumption of longitudinal boost invariance implicit in most hydro approaches.  Csorgo has shown \cite{csorgo} that, by introducing a Hubble-like flow and allowing for a smeared freeze-out temperature, the HBT radii (as well as spectra and elliptic flow) can be well described by a single parameter set.  However, a blast-wave parametrization \cite{retiereLisa} which maintains longitudinal boost invariance also was able to fit the data rather well.  Both of these ``hydro-inspired" approaches do not contain a full hydrodynamic evolution, but the descriptions of freeze-out obtained by their fits possibly hint at the directions full hydro models should take.

A few alternatives to hydro have been investigated.  Studies of the effect of introducing opacity in a parton cascade model \cite{molnar} showed that the pion freeze-out distribution is indeed sensitive to the transport opacity in the partonic phase.  For a parton cascade with no opacity, all three HBT radii show no $k_T$ dependence and values below the data.  As the opacity is increased, $R_{\rm long}$ and $R_{\rm out}$ develop $k_T$ dependences and increase toward the data nicely, but $R_{\rm side}$ doesn't move.  (The authors of Ref.~\cite{molnar} claim this points to a lack of sensitivity of $R_{\rm side}$ to early partonic dynamics.)  Still, the behavior of $R_{\rm long}$ and $R_{\rm out}$ with increasing opacity may indicate that reality lies somewhere between the extremes of cascade and ideal hydro.

Finally, a natural consequence of hydro models is a negative $x$-$t$ correlation, {\it i.e.} pions {\it further} from the source's center are emitted {\it earlier}.  $R_{\rm out}$ depends explicitly on an $\langle xt \rangle$ cross-term, with a negative $\langle xt \rangle$ acting to increase $R_{\rm out}$.  However, cascade models such as AMPT \cite{ampt} have shown that {\it positive} $x$-$t$ correlations arise naturally in their codes, thereby reducing $R_{\rm out}$ compared to hydro's overprediction.  Whether it makes physical sense that particles emitted further from the center could decouple after particles emitted closer to the center is an important question that needs to be confronted by these models.  

\section{Recent advances}

Despite our difficulties to understand what standard pion HBT measurements are telling us at RHIC (and therefore at lower energies, for that matter), two-particle interferometry studies have been extended recently into new domains at both SPS and RHIC.  Here I will discuss two examples.  

STAR has recently completed an analysis of the azimuthal dependence of HBT radii relative to the reaction plane at $\sqrt{s_{NN}}=200$ GeV \cite{starAsHBT}.  Azimuthally sensitive HBT was suggested \cite{asHBT} as a probe of how spatial anisotropy evolves in non-central collisions.  The reasoning is straightforward: (a) The initial almond-shaped geometry gives rise to anisotropies in pressure gradients, the same gradients responsible for elliptic flow.  (b) The pressure gradients drive a preferential expansion in the reaction plane that decreases the spatial anisotropy.  (c) HBT provides a measure of the freeze-out source shape, which in principle could change its orientation from {\it out-of-plane} to {\it in-plane} extended depending on the amount of pressure built up and the expansion time.

\begin{figure}[bt]
\begin{center}	
{\includegraphics*[width=3.5in]{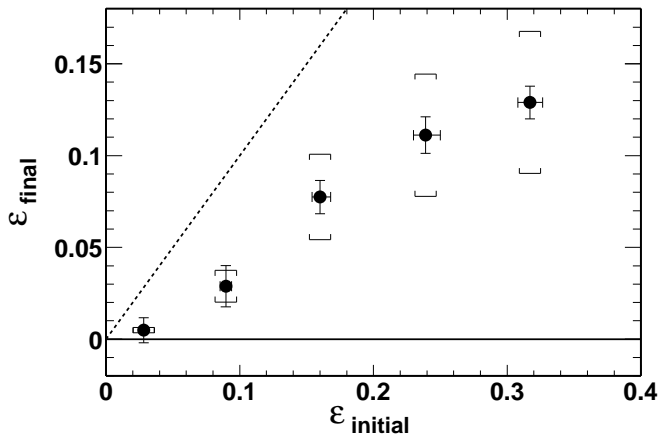}}
\end{center}
  \caption{Source eccentricity obtained with azimuthally-sensitive HBT ($\varepsilon_{\rm final}$) vs.~initial eccentricity
from a Glauber model ($\varepsilon_{\rm initial}$) in Au+Au collisions at $\sqrt{s}=200$ GeV.  The most peripheral collisions correspond to the largest eccentricity. The dashed line indicates $\varepsilon_{\rm initial}$ = $\varepsilon_{\rm final}$.   \cite{starAsHBT}.} \label{fig:two}
\end{figure}

STAR's results showed an intuitive (though model-dependent) centrality dependence of the system's eccentricity at freeze-out (Figure \ref{fig:two}).  
Near-central collisions showed final eccentricities consistent with zero.  When going from central to peripheral collisions, the final eccentricity increased, reflecting the greater initial eccentricity while retaining some of its initial {\it out-of-plane} almond shape.  Given the strong evidence for significant pressure build-up in the system from elliptic flow measurements, the results point to short evolution times as the dominant cause for out-of-plane freeze-out shapes.  

Another domain that holds much promise is HBT of direct photons \cite{peress}.  Unlike pions, $\gamma$-HBT probes the initial state, thereby providing a potential probe of conditions in the deconfined phase.  However it is a major challenge experimentally, due to the low relative direct-$\gamma$ yield and the many sources of pairs with small relative momenta.  To name a few: photon conversions, decays of $\pi^0$'s (which themselves are affected by HBT), resonance decays, misidentified $\gamma$'s.  In spite of the technical challenges, WA98 recently performed \cite{wa98} the first $\gamma$-HBT measurement in heavy-ion collisions, in a low-momentum range of $0.15 < p_T < 0.35$ GeV/$c$.  Very interestingly, they observed HBT radii quantitatively similar to $\pi$-HBT for the same $p_T$, which they attributed to soft photons arising in late stages of the collision.  

Though this was surely already significant enough for a {\it Phys.~Rev.~Lett.}, WA98 took the analysis a step further, using the $\lambda$ parameter from fitting the correlation function (and assuming a fully chaotic source) to determine the absolute direct photon yield in this momentum range.  This technique provides a complementary tool to subtraction-based direct-$\gamma$ yield measurements, and at higher transverse momenta it holds great promise for accessing the temperature reached at early stages of the collision.  (Direct photon studies are underway at RHIC.)

\section{HBT and freeze-out from p+p to Au+Au}

One of the advantages of the RHIC experimental program is the ability to collide different systems with the same center-of-mass energy, allowing for identical analyses of these different systems. Preliminary results were presented at this conference \cite{tom} of pion HBT in p+p and d+Au collisions at $\sqrt{s_{NN}}=200$ GeV, for comparison to the centrality-binned Au+Au analysis (with fixed particle id.~cuts, same $k_T$ bins).  Like the heavy-ion case, the three HBT radii in p+p exhibit a characteristic decrease with increasing $k_T$, though for p+p it has been attributed to string and multistring fragmentation in earlier studies.  A somewhat surprising result comes about when dividing the Au+Au and d+Au radii by the p+p radii \cite{tom}.  The divided trends are roughly flat with $k_T$ for all radii, indicating an apparent scaling in the $k_T$ dependences of the HBT radii for these three systems.  Given that the $k_T$ dependence presumably arises in very different ways (collective expansion in heavy ions, string fragmentation in elementary particles), these results also are a bit puzzling.

\begin{figure}[bt]
	{\includegraphics*[width=5.0in]{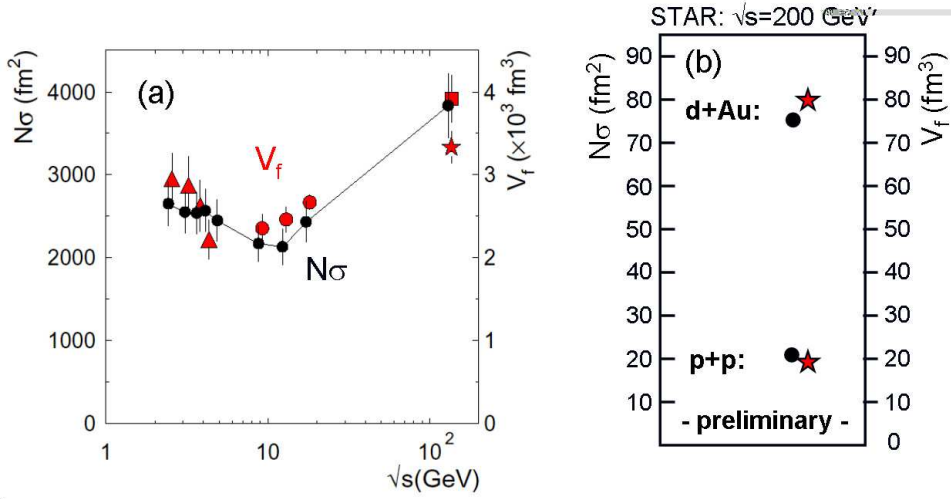}}
  \caption{(a) The freeze-out volume $V_f$ and the density/cross-section term $N\sigma$ as a function of $\sqrt{s}$ for central heavy-ion collisions.  (b) The same quantities calculated for p+p and d+Au at $\sqrt{s}=200$ GeV, plotted on the same relative scale; circles ($\bullet$) are $N\sigma$, stars ($\star$) are $V_f$.}\label{fig:upfo}
\end{figure}

Similarity among these systems can be further tested.  At the last Quark Matter conference, CERES proposed \cite{ceresUPFO} a simple ansatz to investigate if pion freeze-out is driven by a critical mean free path $\lambda_f$.  They developed an approximate expression for $\lambda_f$ that depends on the freeze-out volume $V_f$ estimated from HBT radii, pion and nucleon particle densities ($N_i$), and scattering cross sections for these particles ($\sigma_{ij}$).  Here I only give the final expression for $\lambda_f$; see their paper \cite{ceresUPFO} for full development of, and reasoning behind, this expression and its caveats (such as how the $N_i$ are estimated from the $dN/dy_i$, $R_{\rm long}$, and thermal velocity $\beta_{\rm th}$):
\begin{equation}\label{eq:ceres}
\lambda_f = \frac{V_f}{N\sigma} \equiv 
\frac{(2\pi)^{3/2} R_{\rm long}\, R_{\rm side}^2}{N_N \sigma_{\pi N} + N_\pi \sigma_{\pi\pi}} .
\end{equation}
Figure \ref{fig:upfo}(a) shows the numerator and denominator of Eq.~\ref{eq:ceres} plotted in the same panel vs.~$\sqrt{s}$ (from AGS to RHIC energies), with the scale for $V_f$ ($N\sigma$) defined on the right (left) axis.   
Both trends exhibit non-monotonic behaviors with dips in between AGS and SPS energies that match impressively for the given scales.  $V_f$ decreases at AGS energies mostly due to decreasing $R_{\rm side}$ and increases from SPS to RHIC due to increasing $R_{\rm long}$.  The $N\sigma$ expression first decreases and then increases, reflecting the transition between nucleon and pion dominance as the chemical composition changes with increasing $\sqrt{s}$.  The agreement between the trends yields a critical mean free path of $\sim 1$ fm across the energy range, which the authors attribute to significant opaqueness in the source \cite{ceresUPFO}.

Using the HBT radii for p+p and d+Au presented at this conference and published $dN/dy$ values from STAR \cite{starTOF}, the critical mean free path ansatz can be applied to these systems.  Figure \ref{fig:upfo}(b) shows the incredible result that $\lambda_f$ is $\sim 1$ fm for these light systems as well, even though $V_f$ and $N\sigma$ are lower by two orders of magnitude.  I find this really incredible.  Why would these systems exhibit such similar mean free paths at freeze-out in this simple ansatz?  Is it only a coincidence, for example that $\lambda_f$ is reflecting the considerable opaqueness in Au+Au but merely the system size in p+p and d+Au?  Or has CERES touched upon a deeper connection between the freeze-out volume and freeze-out density that is somehow independent of the actual dynamics in the system?

\section{Final remarks}

The status of HBT interferometry as a signature for the formation of a quark-gluon plasma or the corresponding phase transition remains murky.  This murkiness is due mostly to the persisting disagreement between $\pi$-HBT data and most hydrodynamical models that obtain good agreement with transverse momentum spectra and elliptic flow, as well as HBT's intrinsic sensitivity to the latest stages of the collision.  After discussing the HBT puzzle {\it ad infinitum} with many experts before and during this conference, I am of the opinion that there are primarily two candidate solutions: (a) the growing evidence for a lack of longitudinal boost invariance at RHIC suggests that alternative hydro formulations need to be explored further; (b) parton cascade studies that are able to get close to the HBT data with finite opacity might indicate that, contrary to popular belief, perhaps the hydrodynamic ``limit" has not been reached at RHIC.  Both statements illustrate the importance of HBT in understanding heavy-ion collisions.

However, a number of recent developments have shed light on HBT's ability to unlock information about the nature of the collisions we study.  Azimuthally-sensitive HBT measurements of non-central collisions indicate that the system retains some of its initial almond shape at the end of its evolution, supporting the notion that the system lives shorter than predicted.  First observation of direct photon HBT at SPS energies illustrates the potential for $\gamma$-HBT to probe early conditions in the collision.  Identical experimental conditions and analysis methods for p+p and Au+Au at RHIC provide an important link between these two systems with presumably very different dynamics.  The larger Au+Au dataset taken this year at RHIC also holds the potential for exotic correlation studies between non-identical particles, further and more differential measurements of ``standard" HBT extended to higher transverse momenta, and maybe even studies of the quark gluon plasma using HBT of direct photons.  Though HBT interferometry has been troublesome to many of us in the first years of RHIC running, its future contributions to our understanding of relativistic heavy-ion collisions are assured.

Many people helped in the preparation of this overview.  In particular I would like to thank S.~Bekele, T.~Gutierrez, M.~Heffner, B.~Holzmann, and A.~Kisiel for supplying figures on short notice.  I especially want to thank H.~Appelsh\"auser, T.~Csorgo, U.~Heinz, A.~Kisiel, M.~Lisa, D.~Molnar, S.~Soff, and D.~Teaney for enlightening discussions.  This work was supported by NSF grant PHY--0099476.


\end{document}